\documentclass[twoside]{article}

\usepackage[accepted]{aistats2024}

\usepackage[numbers]{natbib}
\usepackage[utf8]{inputenc} 
\usepackage[T1]{fontenc}    
\usepackage{hyperref}       
\usepackage{url}            
\usepackage{booktabs}       
\usepackage{amsfonts}       
\usepackage{nicefrac}       
\usepackage{microtype}      
\usepackage{graphicx}
\begin{document}

%

%

\twocolumn[

\aistatstitle{ReLExS: Reinforcement Learning Explanations for
Stackelberg No-Regret Learners}

\aistatsauthor{Xiangge Huang \And Jingyuan Li \And  Jiaqing Xie}

\aistatsaddress{ ETH Zurich \And  ETH Zurich \And ETH Zurich } ]

\begin{abstract}
With the constraint of a no regret follower, will the players in a two-player Stackelberg game still reach Stackelberg equilibrium? We first show when the follower strategy is either reward-average or transform-reward-average, the two players can always get the Stackelberg Equilibrium. Then, we extend that the players can achieve the Stackelberg equilibrium in the two-player game under the no regret constraint. Also, we show a strict upper bound of the follower's utility difference between with and without no regret constraint. Moreover, in constant-sum two-player Stackelberg games with non-regret action sequences, we ensure the total optimal utility of the game remains also bounded. 

\end{abstract}

\section{Introduction}
 Stackelberg games \cite{Letch2009Learning, MCKELVEY19956,gametheory} is a potent framework for strategic interactions in economics, security games, and machine learning. These games involve a \textit{leader} who commits to a strategy to optimize its welfare, followed by \textit{followers} who react optimally. Stackelberg Equilibria, the solution concept of these games, is widely studied.

The recent surge of interest in utilizing multi-agent Reinforcement Learning in Stackelberg games underscores the importance of learning Stackelberg games. Many multi-agent system design problems can be modeled as Stackelberg equilibrium issues. Gerstgrasser and Parkes \cite{gerstgrasser2023oracles} offered an approach to this challenge, modeling the learning of Stackelberg Equilibria as a Partially Observed Markov Decision Process (POMOP) and thereby transforming the problem into a multi-agent reinforcement learning task. Nonetheless, research was often confined to specific game types, and we learned a broad class of problems where the follower does not have the no regret characteristic. 

Our project focuses on two-player Stackelberg games with a no regret constraint on follower actions. We first prove that in a looser condition, the two players can always reach the Stackelberg Equilibrium (See Theorem 9). Extending this, we show that the players can consistently achieve the Stackelberg equilibrium in a two-player game under the no regret constraint. We also establish a strict upper bound on the difference in the utility of the follower with and without the no regret constraint. Additionally, in constant-sum two-player Stackelberg games with no regret action sequences, we ensure the total optimal utility of the game remains bounded. Through theoretical analysis and experimental validation, our approach shows that the leader could get the Stackelberg value if the followers are no regret.

\section{Related Work}
\subsection{Learning Stackelberg Equilibria}

Stackelberg equilibria have been rigorously investigated across multiple contexts, encompassing single-shot settings such as normal-form games. Letchford and Conitzer \cite{Letch2009Learning} were among the pioneers of learning a Stackelberg equilibrium, with their research concentrating on single-shot Bayesian games. Their approach entailed solving an optimization problem to ascertain a Stackelberg equilibrium, given an explicit problem description. This methodology was also adopted by Paruchuri et al. \cite{Paruchuri2008Playing}, Xu et al. \cite{Xu2014}, Blum et al. \cite{Blum2014}, and Li et al. \cite{Li2022}.

In more contemporary studies, the focus has been redirected towards Markov games. Zhong et al. \cite{Zhong2021} proposed algorithms to identify Stackelberg equilibria in Markov games. However, their work assumed myopic followers, significantly constraining the applicability of their findings. Brero et al. \cite{brero2021learning} extended this work by employing an inner-outer loop approach, which they dubbed the Stackelberg POMDP. Gerstgrasser and Parkes \cite{gerstgrasser2023oracles} introduced a methodology to frame the learning of Stackelberg equilibria as a POMDP. And they solved the frame using reinforcement learning. Their work offers a solid theoretical underpinning and showcases promising results in matrix games. 

\subsection{Existing Implementations of Stackelberg Equilibrium}

Practical implementations of Stackelberg equilibrium, alongside theoretical analyses, are vital across various game settings. The highway merging game, a real-life two-player zero-sum game, has been shown to have Stackelberg equilibrium \cite{huang2022robust}. Certain machine learning tasks, such as learning the covariance matrix for Wasserstein GANs, learning mixture Gaussians, or training DCGAN on MNIST, could also be modeled as Stackelberg Games \cite{pmlr-v119-fiez20a}. Recent work has addressed computational efficiency in feedback Stackelberg games, such as highway driving, which were solved using KKT conditions and the primal-dual interior point method \cite{li2024computation}.

As highlighted in \cite{gerstgrasser2023oracles}, Stackelberg games with multi-agent RL applied to economic market design, such as formulating optimal tax policies \cite{zheng2022ai}, incentive design \cite{yang2021adaptive}, and incentive shaping \cite{shu2018m, shi2019learning}. Prior works have predominantly focused on iterated matrix games and bilateral trade on Atari 2600, under the framework of PPO with Meta RL \cite{gerstgrasser2023oracles}. These implementations demonstrate the practical significance of Stackelberg equilibrium across a variety of domains, highlighting its versatility and robustness in both theoretical and applied settings.

\subsection{No regret learning}

No regret learning has been extensively examined and studied in the academic literature. The foundational work by \cite{LITTLESTONE1994212} introduced the concept and established its importance in the context of online learning. \cite{FREUND1997119} further developed this theoretical framework through the introduction of the Hedge algorithm, which has become a cornerstone in the field. Their subsequent work \cite{FREUND199979} elaborated and expanded on these ideas, demonstrating the efficacy and robustness of no regret algorithms in various competitive scenarios.

In addition, the study \cite{10.1145/258128.258179} made significant contributions by presenting algorithms that achieve no regret performance in the multi-armed bandit problem, highlighting the practical applications of these techniques in decision-making processes under uncertainty. More recent advancements \cite{deng2019strategizing} have focused on refining these algorithms to improve their efficiency and applicability in complex environments, such as adversarial settings and Markov Decision Processes (MDPs). These developments underscore the versatility and robustness of no regret learning strategies, solidifying their role as essential tools in the broader context of online learning and game theory.


\section{Model and Preliminaries}

In this Section, we establish the foundational background and necessary theorems. Firstly, we define the the formal Markov Game $G$ and the Stackelberg Equilibrium of $G$. Then we introduces the regret measure and the no regret strategy (Definitions 3 and 4), which is a new addition to formal Markov Game $G$. To demonstrate that the leader can always reach the Stackelberg Equilibrium under the no regret learning strategy, we use the reward-average strategy (Definition 7) and prove the leader can reach the Stackelberg Equilibrium under this constraint (Theorem 9).


\subsection{Formal Markov Game Model}

Throughout this paper, we focus on a finite two-player Markov Game $G = (\mathcal{S}, \mathcal{A}, P, R, \gamma, T)$. 



\begin{itemize}
    \item $\mathcal{S} = (s_1, s_2)$ is the \textit{State Space} : a set for all possible agent states.
    \item $\mathcal{A} = (\mathcal{A}_1,\mathcal{A}_2)$ is the \textit{Action Space} : a set for all possible different agent actions.
    \item $P(s'|s, a_1, a_2)$ is the \textit{Transition Function} : describe the probability from state $s$ to state $s'$ after choosing action $a_1,a_2$ for two players. 
    \item $R(s_1, s_2, a_1, a_2)$ is the \textit{Reward Function} : describe the reward got by agent $i$ in state $s_i$ after choosing action $a_1, a_2$ for all two agents. 
    \item $\gamma$ is the \textit{Discount Factor} : describe the factor influencing the future reward. 
    \item $T$ is the \textit{Iteration} : describe the number of iterations that two players play the game. 
\end{itemize}


In this game $G$, we refer to the first player as the \textit{leader} and the second player as the \textit{follower}. And we denote their state-action pair as $(s^L_t, a^L_t)$ and $(s^F_t, a^F_t)$ respectively in round $t$. In each round, the leader will first act, only considering two players' states, meaning that $a_t^L \sim \pi^L(a|s_t^L, s_t^F)$, where $\pi^L(a|s_t^L, s_t^F)$ denotes the leader strategy that maps states to actions. Then the \textit{leader} receives reward $R_t^L$ after the follower takes its action. In contrast, the \textit{follower} can view the leader's action and then choose its action based on this information, meaning that $a_t^F \sim \pi^F(a|s_t^F, a_t^L)$, where $\pi^F(a|s_t^F, a_t^L)$ denotes the follower strategy. Then the \textit{follower} receives reward $R_t^F$.



 Additionally, we denote one round $C_t$ of game $G$ as the joint actions of \textit{leader $(L)$} and \textit{follower $(F)$}, defined as $C_t = (a_t^L, a_t^F)$. The total utility in each round $C_t$ is $U_t(L, F) = R_t^L + R_t^F$. Without loss of generality, we normalize the reward function to let the total utility or reward $|U_t(L, F)| \leq 1$ in each round for simple calculation in this game $G$.




\paragraph{Best Response.} \textit{In this} Markov Game $G$, in round $t$, \textit{after the \textit{leader} chooses a fixed action $\bar{a}_t^L \in \mathcal{A}^L$ in a fixed state $\bar{s}_t^L \in \mathcal{S}^L$ , we call the best response for the \textit{follower} in its current state $s_t^F$
$$a_t^F = \arg \max_{a^F} R_t^F(s_t^F, a^F, \bar{s}_t^L, \bar{a}_t^L)$$}

\textit{The follower's best response means that the follower chooses the action that maximizes its reward or utility based on the limitation of the leader's strategy.}

Based on our previous construction for this game $G$, now we define Stackelberg Equilibrium.

\paragraph{Definition 1.} 
\label{def:1} \textit{The} Stackelberg Equilibrium \textit{of a Markov Game $G$ in round $t$ is a pair of strategies $\pi_t^L(a_t^L|s_t^L)$ and $\pi^F(a_t^F|s_t^F, a_t^L)$ that
maximizes the current leader reward $R_t^L (s_t^L, a_t^L, s_t^F, a_t^F)$ under the constraint that $a_t^F \sim \pi^F(a_t^F|s_t^F, a_t^L)$ is a best-response to $a_t^L \sim \pi^L(a_t^L|s_t^F, s_t^L)$. }


Based on Definition 1, if this game only continues for one round, the leader can choose its best policy, the follower will reach its best response, and the two can always get the Stackelberg Equilibrium. Then we extend this two-player Markov Game to $T$ rounds. For the $T$ rounds for this game, the only thing that changed is that they are supposed to consider all rewards or total utilities they get in the whole $T$ rounds, that is
\begin{equation}
\begin{aligned}
    \mathbb{E}_{d^L(s_0)} \Bigg[ &\sum_{t=0}^T R_t^L(s_t^F, a_t^F, s_t^L, a_t^L) \Bigg| a_t^L \sim \pi^L(a|s_t^F, s_t^L), \\
    & s_0^F \sim d^F(s_0), s_0^L \sim d^L(s_0) \Bigg]
\end{aligned}
\end{equation}
\begin{equation}
\begin{aligned}
    \mathbb{E}_{d^F(s_0)} \Bigg[ &\sum_{t=0}^T R_t^F(s_t^F, a_t^F, \bar{s}_t^L, \bar{a}_t^L) \Bigg| a_t^F \sim \pi^F(a|s_t^F, a_t^L), \\
    & s_0^F \sim d^F(s_0), \bar{s}_t^L, \bar{a}_t^L \Bigg]
\end{aligned}
\end{equation}

For the leader and the follower strategy depends on different factors, we call the leader is non-adaptive, which means its action $a^L$ is only decided by a fixed action space $\mathcal{A}^L$ unaffected by the follower. Conversely, the follower adopts an adaptive strategy, with actions $a_t^F \sim \pi^F(a|s_t, a_t^L)$ and its action space might be influenced $\mathcal{A}^F$ by the action taken by leader $a_t^L$ in round $t$.

\paragraph{Theorem 2.} \textit{Adaptive follower can keep the best response in Markov Game $G$.}

\paragraph{Proof. } \textit{We focus on special round $t$. We know follower is adaptive, so follower action $a_t^F $ can be influenced by previous leader action $a_t^L$. Let $a_t^F = \arg \max_a \pi^F(a|s_t, a_t^L)$, for a fixed leader state $\bar{s}_t^L$ and its corresponding fixed leader action $\bar{a}_t^L$, there exists argmax function to build the function relationship $ f: \bar{s}_t^L, \bar{a}_t^L \rightarrow a_t^F$. In rount $t$, the policy we learned is to maximize the reward expectation, so $a_t^F = \arg \max_a R_t(s_t^F, \pi^F(a_t^F|s_t, a_t^L)) = \arg \max_{a^F} R_t^F(s_t^F, a_t^F, \bar{s}_t^L, \bar{a}_t^L)$, meaning the best response for the follower in round $t$ with leader fixed action $\bar{a}_t^L$ and fixed state $\bar{s}_t^L$ still keeps $a_t^F$ unchanged, and same as other rounds.}

\subsection{no regret Learning and Characteristic}

In our previous Markov game $G$, we established a basic setting involving $T$ rounds of state and action for both the leader and the follower. Each player's objective is to maximize their total reward over all rounds of the game. We begin by defining the regret value for the follower to measure the reward or utility value got by these two players.

\paragraph{Definition 3.} \textit{The regret measure in the follower for two-player game the first $T$ rounds for the action series $\vec{a}^F = (a_1^F, ..., a_T^F)$ is defined as}

\begin{equation}
\begin{aligned}
    & Reg_T(\vec{a}^F) = \max_{\vec{a}^F}  \mathbb{E}_{d^F(s_0)} \Bigg[\sum_{t=0}^T R_t^F(s_t^F, a_t^F, \bar{s}_t^L, \bar{a}_t^L) \Bigg|   \\ a_t^F
    & \sim \pi^F(a|s_t^F, a_t^L),  s_0^F \sim d^F(s_0), \bar{s}_t^L, \bar{a}_t^L \Bigg] - \sum_{t=0}^T \bar{R}_t^F
\end{aligned}
\end{equation}

Here $\bar{R}_t^F$ means the achieved utility or reward got in first $T$ round separately. And this formula describe that the difference between best reward or utility can be got in the first $T$ round and the real value actually acquired by the follower. For any leader action series , we define the best reward operator $\mu_{a_T^F}^*$ for first $T$ rounds

\begin{equation}
\begin{aligned}
    & \mu_{\vec{a}^F}^* R_T^F = \max_{\vec{a}^F}  \mathbb{E}_{d^F(s_0)} \Bigg[\sum_{t=0}^T R_t^F(s_t^F, a_t^F, \bar{s}_t^L, \bar{a}_t^L) \Bigg|  \\ a_t^F
    & \sim \pi^F(a|s_t^F, a_t^L),  s_0^F \sim d^F(s_0), \bar{s}_t^L, \bar{a}_t^L \Bigg]
\end{aligned}
\end{equation}

We define $\mu_{a_T^F}^*$ as the best reward operator, used to calculate the maximal expected reward $R_T^F$ obtainable by the follower in the first $T$ rounds.

\paragraph{Definition 4.} \textit{The follower strategy (or action series $\vec{a}^F$) for Markov Game $G$ has no regret characteristic in first $T$ rounds is defined as}

$$
\mathbb{E} \left[\mu_{\vec{a}^F}^* R_T^F - \sum_{t=0}^T \bar{R}_T^F\right] = o(T)
$$

\textit{where \(o(T)\) denotes a term that grows sublinearly with \(T\), implying that the expected difference between the best possible reward and the accumulated achieved reward per round can not reach the linear reflection as the number of rounds \(T\) increases. }

Based on the regret measure and Theorem 4 in this game $G$, we then consider the follower strategy which might gets slight difference in utility or reward with its best response in current round.

\paragraph{Theorem 5.} \textit{In the first $T$ rounds, if the regret mesure is sublinear, the follower has no regret characteristic for its action series $\vec{a}^F$ and vice versa.}

\paragraph{Proof. } \textit{If the best possible reward exceeds the actual reward by only a sublinear amount, for any action series $\vec{a}^F$ in the first $T$ rounds, it satisfies:
\[
 Reg_T(\vec{a}^F) = \mu_{\vec{a}^F}^* R_T^F - \sum_{t=0}^T \bar{R}_T^F = o(T)
\]
Using the linearity of expectation, we have:
\[
\mathbb{E}[Reg_T(\vec{a}^F)] = \mathbb{E} \left[ \mu_{\vec{a}^F}^* R_T^F - \sum_{t=0}^T \bar{R}_t^F \right] = \sum_{\vec{a}}^F o(T) = o(T)
\]
Thus, the follower have no regert characteristic for its action series $\vec{a}^F$.
}

Then we turn to focus on the leader actions, we apply reinforcement learning algorithm on leader strategy and it can keep no regret strategy for this Markov Game $G$.

\paragraph{Theorem 6.} \textit{Non-adaptive leader can always take strategy to keep no regret characteristic for Markov Game $G$ in first $T$ rounds, which means }
$$
\mathbb{E} \left[\mu_{\vec{a}^L}^* R_T^L - \sum_{t=0}^T \bar{R}_T^L\right] = o(T)
$$
where $\mu_{\vec{a}^L}^* R_T^L$ is the leader's  the best reward operator in first $T$ rounds, which will maximize the sum of leader's reward in first $T$ rounds.

\paragraph{Proof.} 

\textit{Consider a non-adaptive leader with a fixed action space \(a^L \in \mathcal{A}^L\), unaffected by the follower's actions. By assumption, the leader always selects the action that maximizes the accumulated reward in first $T$ rounds. Thus, for the leader, we also have:}

\begin{equation}
\begin{aligned}
    & \mu_{\vec{a}^L}^* R_T^L = \max \mathbb{E}_{d^L(s_0)} \Bigg[\sum_{t=0}^T R_t^L(s_t^F, a_t^F, s_t^L, a_t^L) \Bigg| \\
    & a_t^L  \sim \pi^L(a|s_t^F, s_t^L),  s_0^F \sim d^F(s_0), s_0^L \sim d^L(s_0) \Bigg]
\end{aligned}
\end{equation}

\textit{Similar to Definition 4, we have the definition of the no regret strategy for the leader. Combined the linearity of expectation, we could prove the strategy is no regret}

\begin{equation}
\begin{aligned}
\mathbb{E} \left[ \mu_{\vec{a}^L}^* R_T^L - \sum_{t=0}^T \bar{R}_t^L \right]  &= \mathbb{E} \left[ \mu_{\vec{a}^L}^* R_T^L\right] - \mathbb{E} \left[\sum_{t=0}^T \bar{R_t}^L \right] \\
&= \mu_{\vec{a}^L}^* R_T^L - \mu_{\vec{a}^L}^* R_T^L \\
&= o(T)
\end{aligned}
\end{equation}

\textit{Therefore, a non-adaptive leader keeps no regret characteristic without restirction in its strategy.}

Given Theorem 6, we can disregard the influence of the leader with our reinforcement learning strategy can be viewed as owning no regret characteristic and then focus on the no regret measure for the follower. To measure the influence of different actions taken in the same round for follower utility or reward, we define notation $\mu_{a_T^F}$ as a strategy reward operator, which operates the current reward function $R_T^F$ and gets its corresponding follower reward or utility with current action $a^F_T$ in action series $\vec{a}^F$

\begin{equation}
\begin{aligned}
    &\mu_{\vec{a}^F} R_T^F = \mathbb{E}_{d^F(s_0)} \Bigg[\sum_{t=0}^T R_t^F(s_t^F, a_t^F, \bar{s}_t^L, \bar{a}_t^L) \Bigg| \\
    &  a_t^F \sim \pi^F(a|s_t^F, a_t^L), s_0^F \sim d^F(s_0), \bar{s}_t^L, \bar{a}_t^L \Bigg]
\end{aligned}
\end{equation}

One great characteristic for the action regret measure is that the accumulation of the reward is supposed to keep smooth, without large difference in different action taken in current round $T$. Specifically, we want the reward  rate to be sublinear, as 

$$
\mu_{\vec{a}^F} {R_T}^F - \mu_{\vec{a}^{F'}} {R_T}^F = o(T)
$$

This formula means different action series $\vec{a}^F$ and $\vec{a}^{F'}$ taken in the first  $T$ rounds do not get great influence on the follower cumulative reward or utility, and we define this characteristic as reward-average.

\paragraph{Definition 7.} \textit{The follower strategy is reward-average for any $a_T^F \sim \pi^{F(a|s_T^F, a_T^L)} $ and ${a_T}^{F'} \sim \pi^F(a|s_T^F, a_T^L)$ after $T$ rounds if }

$$\mu_{\vec{a}^F} R_T^F - \mu_{\vec{a}^{F'}} R_T^F = o(T)$$

\paragraph{Theorem 8.} \textit{With reward-average limitation on follower action series $\vec{a}^F$, all follower strategy is be no regret as the best response.}

\paragraph{Proof.} \textit{From Theorem 2, we know an adaptive follower can always maintain the best response in a Markov Game \( G \). Given that the follower's strategy is reward-average for any action series $\vec{a}^F$ and $\vec{a}^{F'}$, it satisfies:}
\[
\mu_{\vec{a}^F} R_T^F - \mu_{\vec{a}^{F'}} R_T^F = o(T)
\]
\textit{Setting \( \vec{a}^F \) as the best reward action series, we have:}
\[
\mu_{\vec{a}^F}^* R_T^F - \mu_{\vec{a}^{F'}} R_T^F = o(T)
\]
\textit{Then,}
\[
\mu_{\vec{a}^F}^* R_T^F - \mathbb{E}_{d^{F'}(s_0)} \left[ \sum_{t=0}^T R_t^F \right] = o(T)
\]
\textit{Using the fact that \( \mu_{\vec{a}^F}^* R_T^F \) is constant and applying the linearity of expectation, we obtain:}
\[
\begin{aligned}
    \mathbb{E}_{d^{F'}(s_0)} \left[ \mu_{\vec{a}^F}^* R_T^F \right] - &\mathbb{E}_{d^{F'}(s_0)} \left[ \sum_{t=0}^T R_t^F \right] \\
    &= \mathbb{E}_{d^{F'}(s_0)} \left[ \mu_{\vec{a}^F}^* R_T^F - \sum_{t=0}^T R_t^F \right] \\
    &= o(T)
\end{aligned}
\]

\textit{For any achieved utility or reward obtained in the first \( t \) rounds, we could satisfy Definition 4:}
\[
\mathbb{E} \left[ \mu_{\vec{a}^F}^* R_T^F - \sum_{t=0}^T \bar{R_t}^F \right] = o(T)
\]
\textit{Thus, with the reward-average constraint on the follower's action series \( \vec{a}^F \), any follower strategy can be no regret as the best response.}

Easy to notice that the reward-average can derive the less different between current reward or utility and optimal reward or utility.

So by Theorem 8, via reward-average strategy, we can view any follower strategy in current round $T$ close to the best response for the leader strategy and do not need to get the actual best response. And we can know that all follower strategy here can also be non-regret.

\paragraph{Theorem 9.} \textit{If the follower strategy in Markov Game $G$ is reward-average, the game can always reach the Stackelberg Equilibrium via the leader taken reinforcement learning and the follower taken no regret algorithm with no regret characteristic.}

\paragraph{Proof Idea.} \textit{From Theorem 8, with reward-average constraints on the follower's action series $\vec{a}^F$, the follower's strategy can achieve no regret and have no regret characteristic as the best response. The follower's best response ensures that it always selects the strategy that maximizes its reward given the leader's strategy constraints. This satisfies Definition 1. Thus, if the follower's strategy in Markov Game $G$ is reward-average, the game can reach Stackelberg Equilibrium with the leader using reinforcement learning and the follower using a no regret algorithm.}

\section{Achieve Restriction Stackelberg Equilibrium}

Based on our previous theorem, we have demonstrated that if the follower's strategy maintains reward-average, the whole game can consistently achieve the Stackelberg Equilibrium under the no regret condition. We now aim to relax the constraints on the follower's strategy and focus on the social reward or utility. Specifically, we aim for the leader and the follower to achieve an average social reward or utility arbitrarily close to the Stackelberg value under no regret conditions.

\paragraph{Theorem 10.} \textit{If the follower's strategy is no regret, then for any $\epsilon > 0$, $|U(L, F) - U_S(L, F)| < \epsilon T + o(T)$ in the first $T$ rounds, where $U_S(D, F)$ is the total utility or reward from the Stackelberg Equilibrium.}

\paragraph{Proof. } \textit{From Theorem 5, if the follower's strategy is no regret, the follower satisfies $$\mu_{\vec{a}^F}^* R_T^F - \sum_{t=0}^T \bar{R}_T^F < \frac{\epsilon}{2}T  + o(T)$$
And by Theorem 6, the leader can keep no regret and in each round $t$, we can get 
$$\mu_{\vec{a}^L}^* R_T^L - \sum_{t=0}^T \bar{R}_T^L <\frac{\epsilon}{2}T  + o(T)$$}
\textit{The difference between the sum of the two-player utility and the total utility from the Stackelberg Equilibrium in the first \(T\) rounds can be expressed as:}

\[
\begin{aligned}
    &\left|U(L, F) - U_{S}(L, F)\right| \\&= \left|\mu_{\vec{a}^F}^* R_T^F - \sum_{t=0}^T \bar{R}_T^F + \mu_{\vec{a}^L}^* R_T^L - \sum_{t=0}^T \bar{R}_T^L\right| \\
    &< \epsilon T + o(T)
\end{aligned}
\]

Thus, according to Theorem 10, the follower can maintain a no regret learning strategy, ensuring that the total social utility or reward of the system does not significantly decrease. Consequently, we establish Theorem 11 as the restricted Stackelberg Equilibrium in our game $G$.

  \paragraph{Theorem 11.}  \textit{If the follower strategy in Markov Game $G$ has no regret characteristic and applied by no regret algorithm, with the leader taking reinforcement learning, the game can still reach the Stackelberg Equilibrium.}

  \paragraph{Proof Idea. } \textit{From Theorem 5, with no regret restriction  constraints on the follower's action series $\vec{a}^F$, we have $|\mu_{\vec{a}^F} R_T^F - \mu_{\vec{a}^{F'}} R_T^F| < \epsilon T + o(T)$. Then for any action series $\vec{a}^F$ and $\vec{a}^{F'}$ with sum of utility $U(L, F)$ and $U'(L, F)$, the difference of utility will satisfy} \[
\begin{aligned}
    &\left|\mu_{\vec{a}^F} R_T^F - \mu_{\vec{a}^{F'}} R_T^F\right| \\  &\leq \left|\mu_{\vec{a}^F} R_T^F - \mu_{\vec{a}^F}^* R_T^F \right| + \left|\mu_{\vec{a}^{F'}} R_T^F - \mu_{\vec{a}^F}^* R_T^F \right| \\
    &< \epsilon T + o(T)
\end{aligned}
\]
   \textit{So we have} $\left|\mu_{\vec{a}^F} R_T^F - \mu_{\vec{a}^{F'}} R_T^F\right|=o(T)$ \textit{, the follower strategy is reward-average (Definition 7). Thus by Theorem 9, the game can reach Stackelberg Equilibrium with the leader using reinforcement learning and the follower using no regret algorithm.}

\section{Experiments}
\subsection{Game Environment}

\paragraph{Environments: Matrix Game.} The matrix game, such as the prisoner's dilemma, serves as a foundational model in Game Theory for constructing equilibriums and understanding strategic decision-making. These games are typically analyzed in a matrix format where each player's strategies and resulting payoffs are outlined. A two-person matrix game is a specific type of matrix game involving two players, each with a finite set of strategies and corresponding payoffs:

\begin{enumerate}
    \item \textbf{Player A's Strategy Set}: finite strategy set $S_A = \{s_{A1}, s_{A2}, \ldots, s_{Am}\}$. 
    
    \item \textbf{Player B's Strategy Set}: finite strategy set $S_B = \{s_{B1}, s_{B2}, \ldots, s_{Bn}\}$. 
    
    \item \textbf{Payoff Functions}: The payoffs for the players are determined by utility functions $u_A(s_A, s_B) \in \mathbb{R}$ and $u_B(s_A, s_B) \in \mathbb{R}$ where $\mathbb{R}$ is the set of real numbers. These payoffs depend on the strategy combination or strategy profile $s = (s_A, s_B) \in S_A \times S_B$. 

    \item \textbf{Game Squence}: The action taken sequence for the players are non-simultaneous, that is the player A will choose an action firstly and player B then choose an action. In this term, the two player reach state $s_A$ and $s_B$ and get reward $u_A$ and $u_B$.
\end{enumerate}

Especially, we choose 12 iterated matrix games \cite{gerstgrasser2023oracles}. These matrix games are \textit{Markov Games} as played iteratively by two players A and B in a sequence of stages. And we list the game setting in Appendix.
\subsection{Experiment Results}

We test our no regret settings across 12 specific matrix games, examining the dynamics between oracles and followers \cite{gerstgrasser2023oracles} with same setting except restriction in follower. We estimated both original regret and our no regret settings where every environment has shown a Stackelberg Equilibrium, with the leader taking normal reinforcement learning and the follower taking no regret learning algorithm applied by no regret characteristic.

\begin{figure*}[htbp]
    \centering
    \includegraphics[width=\linewidth]{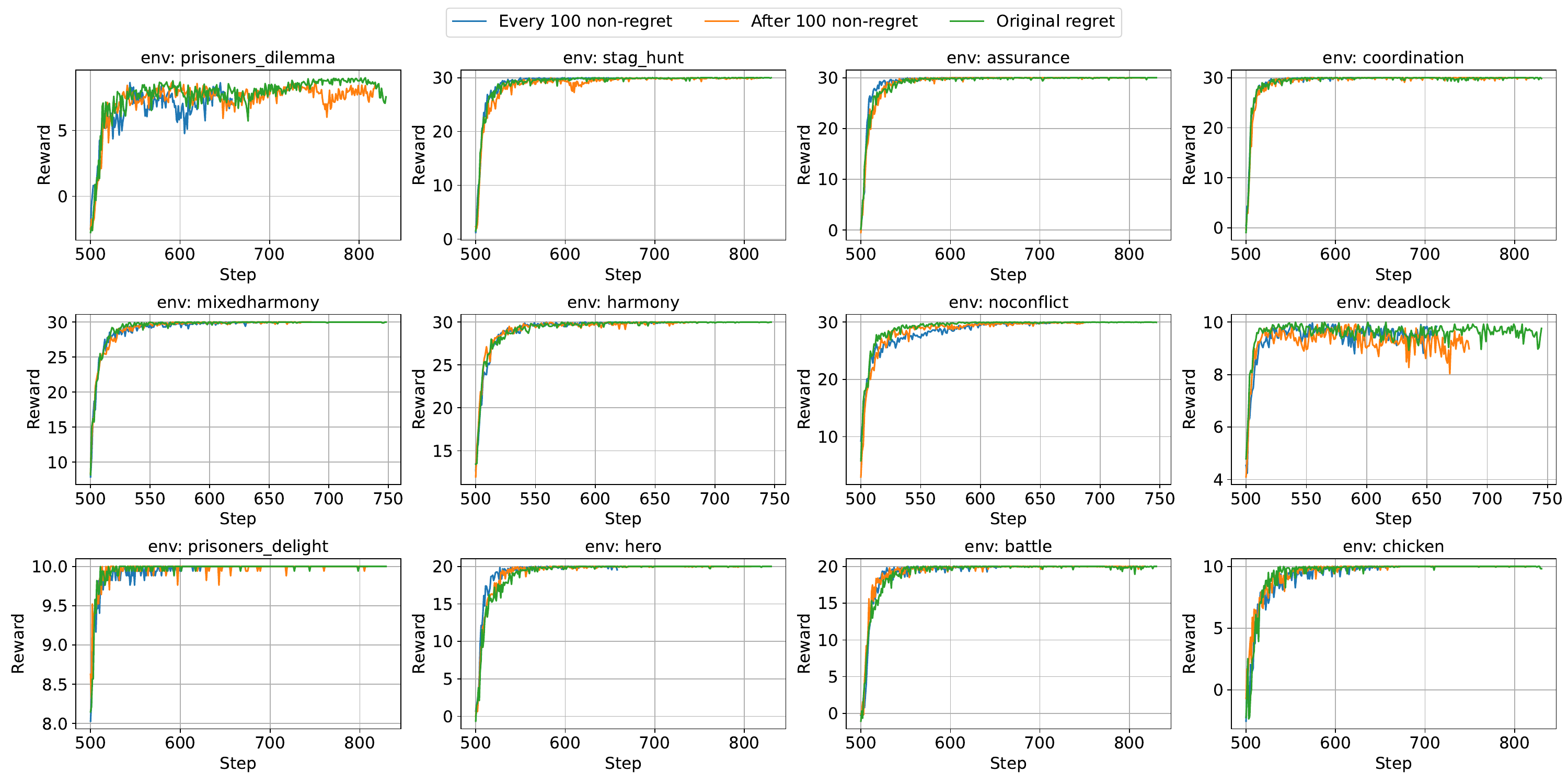}
    \caption {Mean episode reward of PPO+Meta+No Regret RL on 12 canonical symmetric iterated matrix games followed by oracles and followers \cite{gerstgrasser2023oracles}. Green: original regret setting. Blue: no regret every 100 epochs. Orange: no regret after 100 epochs.}
    \label{fig:full_matrices}
\end{figure*}

\begin{figure*} [!htb]
    \centering
    \includegraphics[width=\linewidth]{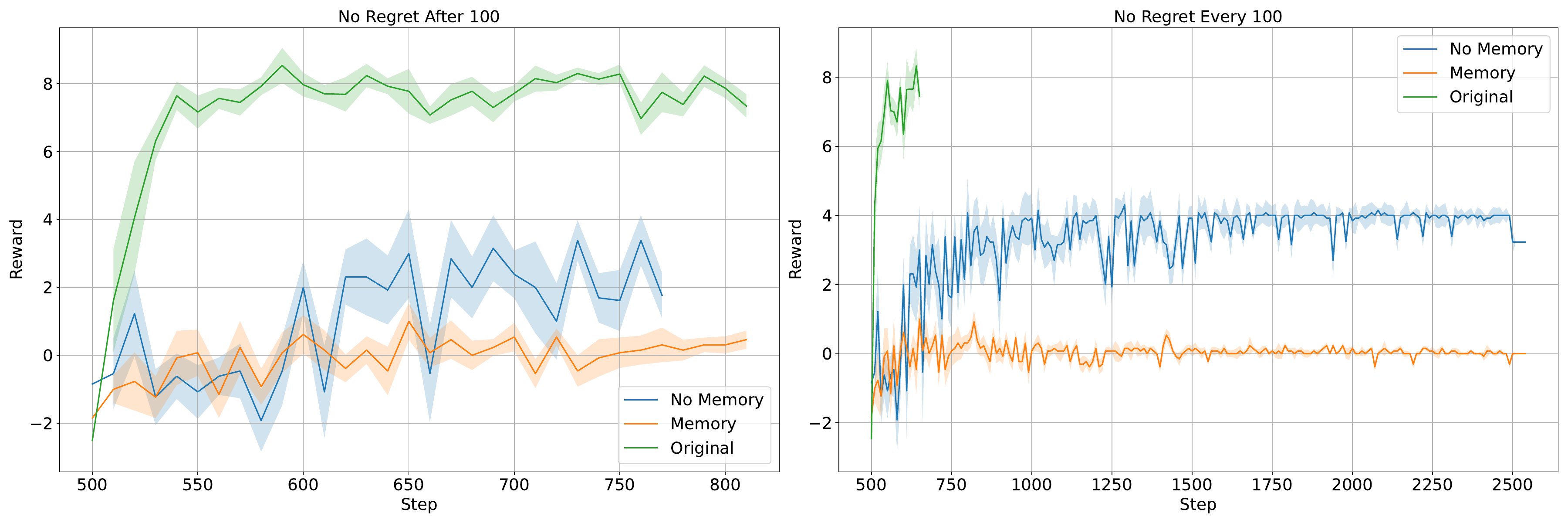}
    \caption{Memory to Leaders. Env: Prisoner's Dilemma}
    \label{fig:memory}
\end{figure*}

The Figure \ref{fig:full_matrices} shows all three strategies could lead to the same reward for all environments except for prisoners\_dilemma and deadlock where our no regret settings perform little worse than regret settings but most of them are on-par. Additionally, for the usage of memory to leader, we found a pattern similar to the regret settings, showing adding memory could break things where reward is low \ref{fig:memory}.

\section{Conclusion}
In conclusion, our research presents a comprehensive framework for achieving Stackelberg Equilibria under no regret constraints in Markov Games using multi-agent reinforcement learning. Building on the work of Gerstgrasser and Parkes \cite{gerstgrasser2023oracles}, we expand their methodology by incorporating the crucial no regret feature. Our theoretical and empirical analyses confirm that integrating no regret constraints into followers' actions facilitates the realization of Stackelberg Equilibria, providing foundations for strategic interactions in various domains such as economics, security games, and machine learning.

We have demonstrated that under mild assumptions, two players can consistently achieve Stackelberg equilibrium. We also established a strict upper bound on the utility difference between the actual game strategy and the Stackelberg strategy, ensuring bounded optimal utility in constant-sum two-player games with no regret action sequences. This work lays the groundwork for future research to refine these models, explore more complex environments, and validate their practical applications, advancing the capabilities and understanding of multi-agent reinforcement learning in game theory.

Future work will focus on refining and expanding our reinforcement learning model for continuous Stackelberg games by exploring more complex and dynamic environments. We aim to apply our approach to multi-agent settings beyond two-player scenarios and in diverse domains such as economics, cybersecurity, and autonomous systems. Additionally, we plan to enhance the computational efficiency of our algorithms and incorporate more sophisticated no regret learning techniques to handle highly adversarial and stochastic environments. Conducting extensive empirical studies to validate our theoretical findings in real-world applications will be essential to demonstrate the practical utility and scalability of our proposed methodologies.


\medskip
\bibliographystyle{apalike}
\bibliography{ref.bib}


\clearpage
\appendix
\section{Appendix}
\subsection{Matrix Game Setting}
\label{why_stackelberg}

In the context of Oracles \& Followers\cite{gerstgrasser2023oracles},  iterated games are considered Stackelberg games due to the inherent leader-follower dynamic that characterizes these games. We reuse its 12 matrix games as Stackelberg games to train the leader in reinforcement learning and the follower in no regret learning algorithm, with additional no regret characteristic applied.

\paragraph{Oracle and Follower Dynamics}
 Oracles \& Followers \cite{gerstgrasser2023oracles} introduces the concept of an oracle to determine the best responses of the followers to the leader’s strategy. This oracle-based approach further solidifies the leader-follower structure by explicitly modeling how followers adapt to the leader’s strategies, a key aspect of Stackelberg games.

\paragraph{Examples and Applications}
In experiment, we provides examples such as the Iterated Prisoner’s Dilemma, where a leader might commit to a Tit-for-Tat strategy, influencing the follower to cooperate. Such examples demonstrate the practical application of Stackelberg equilibria in iterated settings, and we list these iterated matrix games in Table \ref{table:payoff_matrices}.

\begin{table}[h!]
\centering
\renewcommand{\arraystretch}{1.5} 
\begin{tabular}{ccc}
\hline
\multicolumn{3}{c}{\textbf{Iterated Matrix Games}} \\ 
\hline
\textbf{Name} & \textbf{Leader Payoff} & \textbf{Follower Payoff} \\ 
\hline
prisoners dilemma & $\begin{pmatrix} -1 & -3 \\ 0 & -2 \end{pmatrix}$ & $\begin{pmatrix} -1 & 0 \\ -3 & -2\end{pmatrix}$ \\
\hline
stag hunt & $\begin{pmatrix} 0 & -3 \\ -1 & -2 \end{pmatrix}$ & $\begin{pmatrix} 0 & -1 \\ -3 & -2 \end{pmatrix}$ \\
\hline
assurance & $\begin{pmatrix} 1 & -2 \\ 0 & -1 \end{pmatrix}$ & $\begin{pmatrix} 0 & -1 \\ -2 & -3 \end{pmatrix}$ \\
\hline
coordination & $\begin{pmatrix} 0 & -2 \\ 0 & -3 \end{pmatrix}$ & $\begin{pmatrix} 0 & -3 \\ -2 & -3 \end{pmatrix}$ \\
\hline
mixedharmony & $\begin{pmatrix} 0 & -1 \\ -1 & -3\end{pmatrix}$ & $\begin{pmatrix} 0 & -3 \\ -1 & -3 \end{pmatrix}$ \\
\hline
harmony & $\begin{pmatrix} 0 & -1 \\ -2 & -3 \end{pmatrix}$ & $\begin{pmatrix} 0 & -2 \\ -1 & -3 \end{pmatrix}$ \\
\hline
noconflict & $\begin{pmatrix} 0 & -2 \\ -1 & -3 \end{pmatrix}$ & $\begin{pmatrix} -1 & -3 \\ 0 & -2 \end{pmatrix}$ \\
\hline
deadlock & $\begin{pmatrix} -2 & -3 \\ -1 & 0 \end{pmatrix}$ & $\begin{pmatrix} -2 & 0 \\ -3 & -1 \end{pmatrix}$ \\
\hline
prisoners delight & $\begin{pmatrix} 0 & -2 \\ -1 & -3 \end{pmatrix}$ & $\begin{pmatrix} 0 & -3 \\ -2 & -1 \end{pmatrix}$ \\
\hline
hero & $\begin{pmatrix} 0 & -3 \\ -2 & -1 \end{pmatrix}$ & $\begin{pmatrix} -3 & -1 \\ 0 & -2 \end{pmatrix}$ \\
\hline
battle & $\begin{pmatrix} -1 & -2 \\ -2 & -3 \end{pmatrix}$ & $\begin{pmatrix} -2 & -3 \\ -1 & 0 \end{pmatrix}$ \\
\hline
chicken & $\begin{pmatrix} -1 & -2 \\ 0 & -3 \end{pmatrix}$ & $\begin{pmatrix} -1 & 0 \\ -2 & -3 \end{pmatrix}$ \\
\hline
\end{tabular}
\vspace{4pt}
\caption{ Payoff Matrices used in the matrix-game experiments}
\label{table:payoff_matrices}
\end{table}

\newpage
\subsection{Experiment Outcome for Memory and Non-Memory}

\subsection{Experiment Outcome for Regret hidden and No Regret hidden}
\begin{figure*} [!htb]
    \centering
    \includegraphics[width=\linewidth]{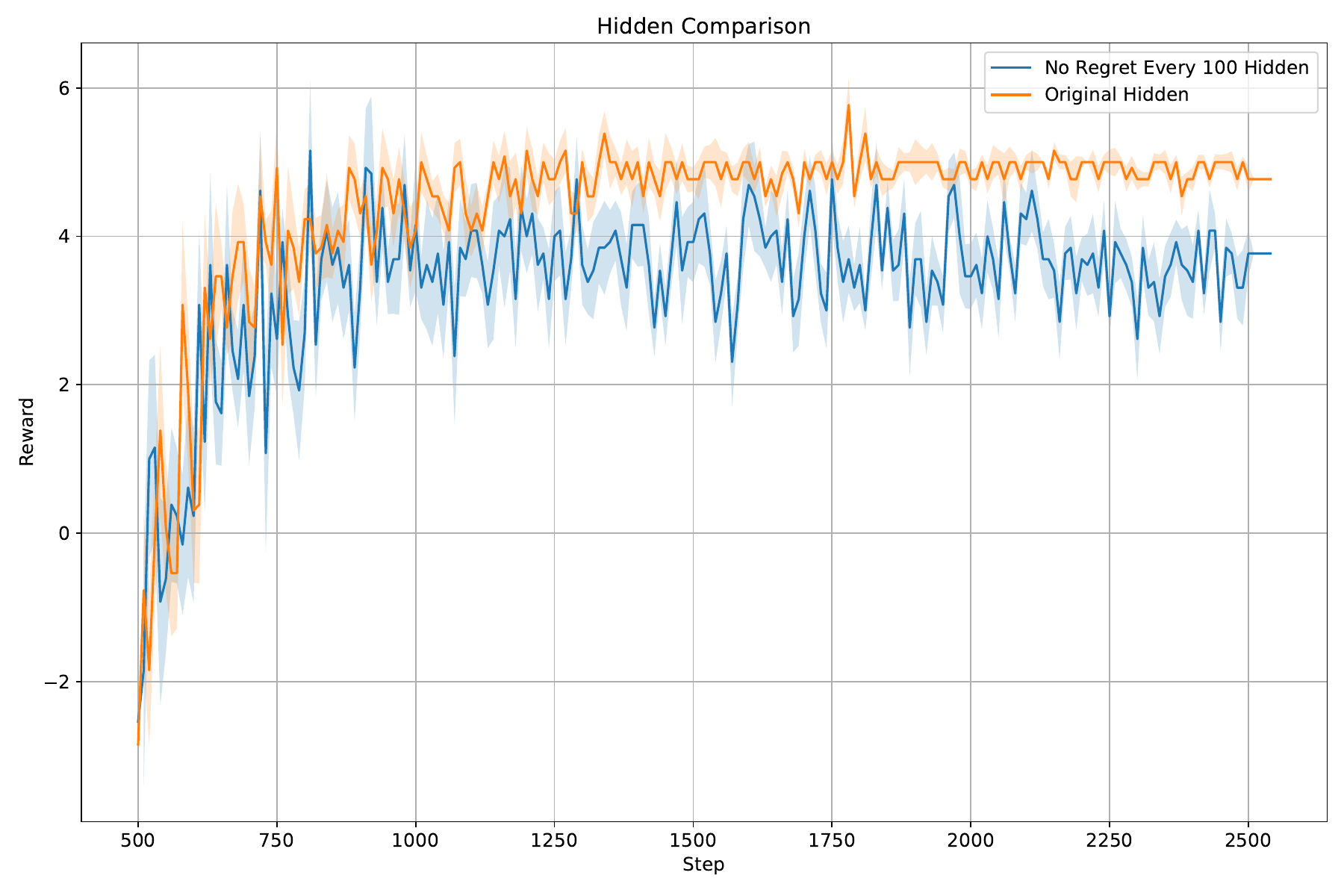}
    \caption{Original Regret Hidden vs. No Regret Hidden}
\end{figure*}

\end{document}